\documentclass[final]{svjour3}
\usepackage{graphicx}
\usepackage{amssymb}
\usepackage{mathptmx}
\usepackage[super]{natbib}
\usepackage{color}
\usepackage{textcomp}
\makeatletter
\journalname{Journal of Low Temperature Physics}
\bibpunct{}{}{,}{s}{}{,}

\begin{document}

\newcommand{\hdblarrow}{H\makebox[0.9ex][l]{$\downdownarrows$}-}
\title{Microstructure analysis of bismuth absorbers for transition-edge sensor X-ray microcalorimeters}

\author{Daikang Yan$^{1,2}$ \and  Ralu Divan$^{1}$ \and Lisa M. Gades$^{1}$ \and Peter Kenesei$^{1}$ \and Timothy J. Madden$^{1}$ \and Antonino Miceli$^{1}$ \and Jun-Sang Park$^{1}$ \and  Umeshkumar M. Patel$^{1}$ \and  Orlando Quaranta$^{1,2}$ \and  Hemant Sharma$^{1}$ \and  Douglas A. Bennett$^{3}$ \and  William B. Doriese$^{3}$ \and  Joseph W. Fowler$^{3}$ \and  Johnathon Gard$^{3,4}$ \and  James Hays-Wehle$^{3}$ \and Kelsey M. Morgan$^{3,4}$ \and Daniel R. Schmidt$^{3}$ \and Daniel S. Swetz$^{3}$ \and Joel N. Ullom$^{3,4}$}

\institute{
1 Argonne National Laboratory, Argonne, Illinois 60439, USA \\
2 Northwestern Univeristy, Evanston, Illinois 60208, USA\\
3 National Institute of Standards and Technology, Boulder, Colorado 80305, USA \\
4 University of Colorado, Boulder, Colorado 80309, USA \\
\email{amiceli@anl.gov}}

\maketitle

\begin{abstract}

Transition-edge sensors (TESs) as microcalorimeters offer high resolving power, owning to their sharp response and low operating temperature. In the hard X-ray regime and above, the demand for high quantum-efficiency requires the use of absorbers. Bismuth (Bi), owing to its low heat carrier density and high X-ray stopping power, has been widely used as an absorber material for TESs. However, distinct spectral responses have been observed for Bi absorbers deposited via evaporation versus electroplating. Evaporated Bi absorbers are widely observed to have a non-Gaussian tail on the low energy side of measured spectra. In this study, we fabricated Bi absorbers via these two methods, and performed microstructure analysis using scanning electron microscopy (SEM) and X-ray diffraction microscopy. The two types of material showed the same crystallographic structure, but the grain size of the evaporated Bi was about 40 times smaller than that of the electroplated Bi. This distinction in grain size is likely to be the cause of their different spectral responses. 

\keywords{Transition-edge sensors, Bismuth, Electroplating, X-ray diffraction}

\end{abstract}

\section{Introduction}

Bismuth (Bi) is a semimetal characterized by small heat capacity and large atomic number, making it a good material for X-ray absorbers for transition-edge sensor (TES) microcalorimeters. Previous studies have shown eV-level energy resolution achieved by TESs with such absorbers [\citenum{Bandler:2008fv,Brown:2008fh,Tatsuno:2016el,Doriese:2017kr}]. However, when deposited via different methods, Bi absorbers exhibit different spectral response. In particular, evaporated Bi (evap-Bi) absorbers commonly exhibit a tail on the low energy side of the spectra, complicating the analysis of the energy peaks. Conversely, electroplated Bi (elp-Bi) absorbers do not exhibit this problem. In Ref.~\citenum{Yan:2017we}, we link this difference to the different grain size distributions in the two types of Bi absorbers. In this work, we present microstructure analysis of the two types of Bi absorbers using scanning electron microscopy (SEM), and high energy X-ray diffraction measurements. 

\section{Fabrication} 

The two samples of Bi (evap-Bi and elp-Bi) for our study were deposited under the similar condition used in the real device fabrication [\citenum{Yan:2017we}]. For the evap-Bi sample, an 80 nm gold (Au) seed layer was e-beam evaporated on a 5 nm titanium (Ti) adhesion layer, and then a 3 \textmu m Bi layer was thermally evaporated at a rate of 100 \AA/s at room temperature on a silicon wafer. The 3 \textmu m elp-Bi layer was electrodeposited on a 1 \textmu m Au seed layer on top of the silicon wafer [\citenum{Gades:2017ix}]. The Bi plating rate was 283 nm/min with a current density of 6.0 mA/cm$^{2}$ DC and a bias voltage of 1.4 V; plating solution was at room temperature and not agitated.

\section{Photon absorption} 

In a TES microcalorimeter, when a high-energy X-ray photon hits the absorber, it liberates an electron (i.e., a photoelectron) from the bonding atom, attributing a kinetic energy that equals the difference between incident photon energy and the work function. While traveling through the material, the photoelectron generates secondary electrons and phonons, which diffuse in the absorber, generating a complex energy distribution between the electron and phonon system and thermalizing the absorber to an average temperature higher than the thermal bath. This energy then flows to the TES, creating a current signal proportional to the energy of the incident photon. Finally, the energy is drained to the heat sink through the thermal connection (e.g., a perforated SiN membrane) between the absorber-sensor and the heat sink, bringing the system back to equilibrium. 

During the thermalization process, if some of the heat carriers were trapped in a band gap or in grain boundaries, the measured energy would be smaller than the real value, smearing the energy spectra to the low-energy side [\citenum{Sun:2015fr}]. Several groups have observed this phenomenon in TESs that have evap-Bi absorbers . It has also been reported that the fraction of energy smeared to the low-energy tail increases with the absorber thickness [\citenum{Doriese:2017kr}], and with the incident X-ray energy [\citenum{Tatsuno:2016el, Yan:2017we}].

The incident photon energy may influence the energy trapping probability because photons with higher energies create larger secondary electron clouds within a material, thus increasing the energy trapping probability. Fig. 1 shows the size of the secondary electron cloud in bulk Bi as a function of incident photon energy computed using the formulation presented in Tabata et al. [\citenum{Tabata:1996tt}]. The Bi mean excitation energy (823 eV) was found in Ref.~\citenum{Seltzer:1984to}. Based on this calculation, the low-energy tail response is more likely to be present if grain size of Bi is on the order of tens of nanometers (i.e., smaller than or comparable to the secondary electron cloud). 

\begin{figure}[htbp]
\begin{center}
\includegraphics[width=0.6\linewidth, keepaspectratio]{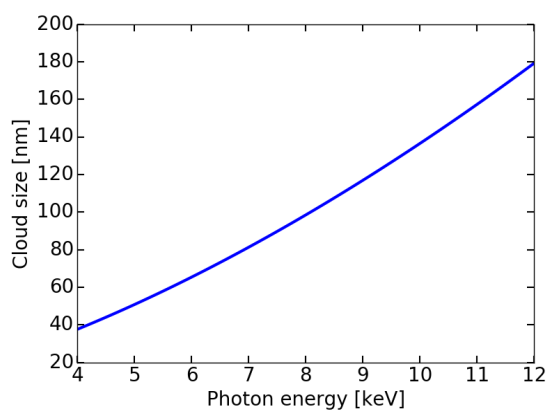}
\caption{Secondary electron cloud size for bulk Bi as a function of incident photon energy. The mean excitation energy used for calculation was 823 eV.}
\end{center}
\label{fig:SE_cloud}
\end{figure}

\section{Characterization}

To characterize the Bi microstructure, we took a series of SEM images of the two types of Bi, examples of which are shown in Fig. 2. The films present a very different appearance, with the elp-Bi grains appearing significantly larger than those of the evap-Bi. 

The SEM images show that the elp-Bi sample has a very rough surface with grain size on the order of \texttildelow 1 \textmu m. On the other hand, evap-Bi shows a much finer surface, with grains that are on the order of \texttildelow 100 nm in size. 

\begin{figure}[htbp]
\begin{center}
\includegraphics[width=1\linewidth, keepaspectratio]{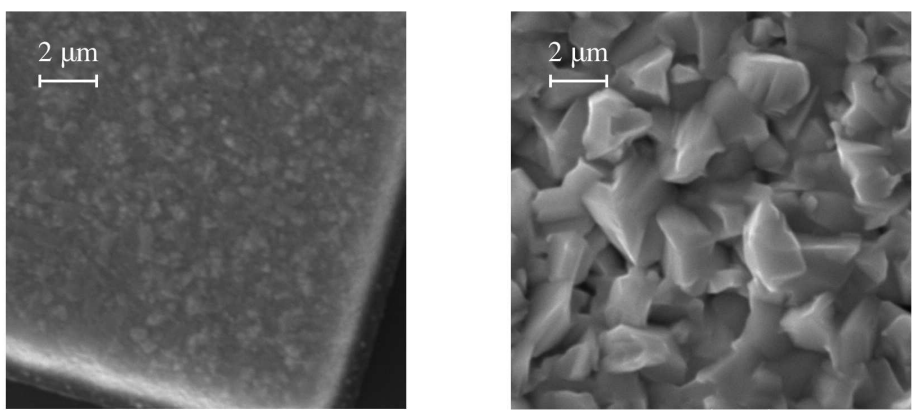}
\caption{SEM images of thermally evaporated Bi (\textit{Left}) and electroplated Bi (\textit{Right}). The images were taken under the same magnification. The electroplated sample clearly shows larger grains than the evaporated sample.}
\end{center}
\label{fig:SEM}
\end{figure}

Although the differences in the two materials are quite evident from these images, a more quantitative analysis is needed to evaluate how the difference in microstructure could influence the absorbers’ performances under X-ray illumination. We therefore characterized the films using high energy X-ray diffraction at the 1-ID-E beamline of the Advanced Photon Source, Argonne National Laboratory. Two measurement techniques were employed, as the two samples have very different microstructure. For the elp-Bi sample, the far field high energy diffraction microscopy (FF-HEDM) technique was used. FF-HEDM is an extension of the rotating crystal method; diffraction spots from large coherent crystals in a polycrystalline sample like the elp-Bi sample are recorded on an area detector as the sample is rotated with respect to the incident monochromatic X-ray beam. Based on the number of diffraction spots recorded for a particular family of crystallographic planes and the size of the illuminated volume, the average size of grains can be computed. 

Fig. 3 illustrates the FF-HEDM setup used in this work. A monochromatic X-ray beam (wavelength $\lambda$ = 0.015358 nm, energy $E$ = 80 keV) was used. To minimize scattering from the substrate, the beam was vertically focused to 1 \textmu m using a set of sawtooth lenses [\citenum{Said.2010yg}] and horizontally cut to 100 \textmu m using a set of slits. The diffraction patterns were recorded on an amorphous-Si detector [\citenum{Lee:2008hv}] placed \texttildelow 1 m from the sample. The sample was rotated along its plane normal direction by 360\textdegree, and diffraction patterns were recorded at each 0.25\textdegree step. 

For the evap-Bi sample, the wide angle X-ray scattering (WAXS) technique was used. WAXS uses the same setup as the one used for FF-HEDM, and records powder diffraction patterns from small grains like the evap-Bi sample on an area detector. For both FF-HEDM data (elp-Bi) and WAXS data (evap-Bi), the diffraction patterns matched the monoclinic Bi phase (symmetry group C2/m with lattice parameters a = 7.8873Å, b = 4.5572Å, c = 6.5836 Å, $\alpha{α}$ = $\gamma$ = 90\textdegree, $\beta$ = 143\textdegree) described by Shu et al. [\citenum{Shu:2016hh}].

\begin{figure}[htbp]
\begin{center}
\includegraphics[width=0.7\linewidth, keepaspectratio]{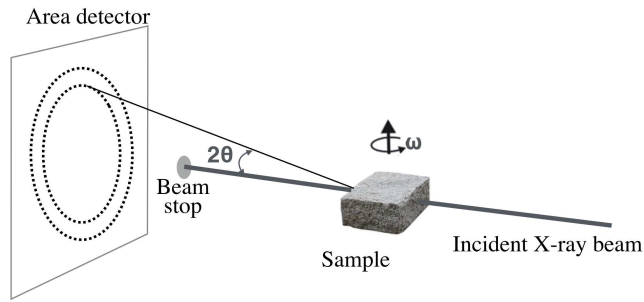}
\caption{Far-field high-energy diffraction microscopy (FF-HEDM) setup. Transmission diffraction patterns were collected at each 0.25\textdegree step during the rotation in the plane normal directionω.}
\end{center}
\label{fig:setup}
\end{figure}

\begin{figure}[htbp]
\begin{center}
\includegraphics[width=1\linewidth, keepaspectratio]{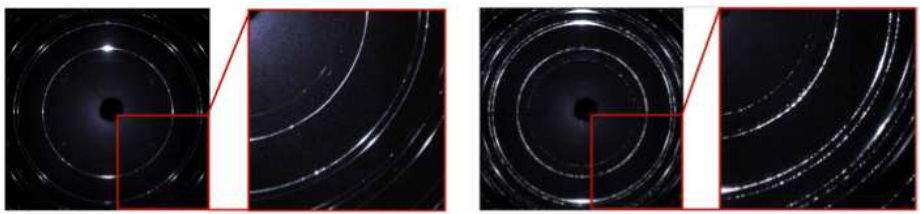}
\caption{Example WAXS and FF-HEDM diffraction patterns for evap-Bi (\textit{Left} and insert) and elp-Bi (\textit{Right} and insert), respectively. The monoclinic Bi phase is the major constituent in both samples but the diffraction patterns are significantly different; the diffraction pattern from evap-Bi is more continuous around the azimuth while that from elp-Bi is spottier.}
\end{center}
\label{fig:XRD}
\end{figure}

The diffraction patterns from the evap-Bi and elp-Bi samples are shown in Fig. 4 \textit{Left} and \textit{Right}, respectively. While the monoclinic phase of Bi is the major constituent in both samples, the diffraction patterns from the two samples show significant differences. The rings are continuous and their intensities are more uniform around the azimuth for the evap-Bi sample (Fig. 4 \textit{Left}); the rings are spotty and the intensities vary significantly around the azimuth for elp-Bi sample (Fig. 4 \textit{Right}). We attribute this observation to different grain sizes as indicated by the SEM images.

For the diffraction patterns from the elp-Bi sample, the total number of diffraction spots for a particular family of crystallographic plane was determined using the appropriate spot searching algorithm [\citenum{MIDAS}]. Given its multiplicity, the total number of grains in the illuminated volume can be estimated and the average size of the constituent grains can be computed. Here, the diffraction spots associated with the \{001\} family of crystallographic planes (multicity of 2) were used. Assuming that the grains are spherical, the diameter of Bi grains in the elp-Bi sample was approximately 1.4 \textmu m.

For the diffraction patterns from evap-Bi sample, the peak widths and Scherrer equation $B = K\lambda /(L cos \theta)$ are used to estimate the average size of the grains. In the Scherrer equation, $B$ is the width of the diffraction peak obtained by fitting the diffraction peak with a pseudo-Voigt function, $K$ is the shape factor (0.93), $L$ is the average size of the grains, and $\theta$ is the Bragg angle associated with a particular diffraction peak. The average grain size in the evap-Bi sample based on this approach is approximately 30 nm. In the raw diffraction pattern, the regions with bright diffraction spots emanating from the substrate were avoided.

These analyses confirmed the qualitative results obtained via the SEM imaging, that there is a sizeable difference in the morphology of the two Bi samples. For grain dimensions below \texttildelow 50 nm, Bi can be of semiconductive nature [\citenum{Hoffman1993,Dresselhaus:1998bg,Lin:2000cj,Lin:2000bg}]. Moreover, given the secondary electron cloud sizes in Fig. 1, energy carriers in the evap-Bi could easily encounter grain boundaries and get trapped. Conversely, the elp-Bi, characterized by grains of \texttildelow 1.4 \textmu m on average, are semi-metallic in nature, and are bigger than the typical secondary electron cloud, greatly reducing the chances of energy trapping at grain boundaries.

\section{Conclusions}

In summary, we have fabricated Bi absorbers via thermal evaporation and electroplating, and characterized them with SEM and high energy X-ray diffraction measurements (WAXS and FF-HEDM). The SEM showed distinct grain sizes in the two absorbers. The high energy X-ray diffraction measurements confirmed this observation and allowed us to quantify the average grain size for both films: \texttildelow 30 nm for evap-Bi and \texttildelow 1.4 \textmu m for elp-Bi. These results support the hypothesis that the different response under X-ray illumination when used for TES microcalorimeters could be due to the very different microstructure of the two films. In particular, the average grain size seems to play a crucial role in the thermalization processes and consequent X-ray energy measurement. 

\begin{acknowledgements}
This work was supported by the Accelerator and Detector R\&D program in Basic Energy Sciences’ Scientific User Facilities (SUF) Division at the Department of Energy. This research used resources of the Advanced Photon Source and Center for Nanoscale Materials, U.S. Department of Energy Office of Science User Facilities operated for the DOE Office of Science by the Argonne National Laboratory under Contract No. DE-AC02- 06CH11357. The contribution of NIST is not subject to copyright. 
\end{acknowledgements}

\pagebreak

\bibliographystyle{JLTPv2}
\bibliography{Bi_XRD_ltd17}

\end{document}